\documentclass[fleqn,10pt, colorlinks=true,allcolors=blue,colorlinks=true,linkcolor=blue,citecolor=blue,urlcolor=blue]{wlscirep}
\usepackage[T1]{fontenc}
\usepackage{graphicx}
\usepackage{bm}

\usepackage{soul}
\usepackage{orcidlink}
\usepackage{amsmath}
\usepackage{amssymb}
\usepackage{enumerate}

\usepackage{xspace}
\usepackage{mathrsfs}
\usepackage{mathptmx}
\usepackage[utf8]{inputenc}
\usepackage{physics}
\usepackage{ulem}
\usepackage{verbatim}

\newcommand{\Neel}{N\'eel\xspace}

\begin{document}

\author[1]{Jackson~L.~Ross}
\author[1,2]{Sarah~Jenkins}
\author[3,4]{Ricardo~Rama-Eiroa}
\author[5,6]{Paul-Iulian~Gavriloaea}
\author[1]{Jack~Collings}
\author[5]{Unai~Atxitia}
\author[5]{Oksana~Chubykalo-Fesenko}
\author[2]{Karin Everschor-Sitte}
\author[7,8]{Rub\'en M.~Otxoa}
\author[1]{Roy~W.~Chantrell}
\author[1]{Richard~F.~L.~Evans}

\affil[1]{School of Physics, Engineering and Technology, University of York, York, YO10 5DD, United Kingdom}
\affil[2]{TWIST Group,
Faculty of Physics and Center for Nanointegration Duisburg-Essen (CENIDE), University of Duisburg-Essen, 47057 Duisburg, Germany
}
\affil[3]{Institute for Condensed Matter Physics and Complex Systems, School of Physics and Astronomy, The University of Edinburgh, Edinburgh, United Kingdom}
\affil[4]{Higgs Centre for Theoretical Physics, The University of Edinburgh, Edinburgh, United Kingdom}
\affil[5]{Instituto de Ciencia de Materiales de Madrid, CSIC, Cantoblanco, 28049, Madrid, Spain}
\affil[6]{National Institute of Research and Development for Technical Physics, 47 Mangeron Boulevard, 700050 Iasi, Romania}
\affil[7]{Hitachi Cambridge Laboratory, J. J. Thomson Avenue, Cambridge CB3 0HE, United Kingdom}
\affil[8]{Donostia International Physics Center, 20018 San Sebasti\'an, Spain}

\title{Temperature-dependent magnetic properties of the antiferromagnet Mn$_\mathbf{2}$Au}

\begin{abstract}
Mn$_2$Au is one of the most promising materials for antiferromagnetic spintronics due to its high \Neel temperature, moderate anisotropy, layered spin structure allowing current induced spin-orbit-torques, and ultrafast spin dynamics. Further understanding and control of the magnetic order in Mn$_2$Au for ultrafast spintronics applications via ultrashort electric/optical means requires accurate understanding of how its magnetic properties depend on temperature. Here we present atomistic spin dynamics simulations of the temperature-dependent macroscopic magnetic properties of Mn$_2$Au single crystals. We introduce a spin Hamiltonian based on previous \textit{ab initio} calculations which produces temperature-dependent scaling of the anisotropy constants with magnetisation consistent with Zener-Akulov-Callen-Callen power laws.
The calculated  temperature dependence of exchange stiffness follows a scaling law with magnetisation  of $m_e^{1.83}$, and predicts a temperature-dependent domain wall width that agrees very well with direct simulation. Our model provides an understanding of temperature-dependent magnetic properties of Mn$_2$Au that enables large-scale simulation of the working conditions of spintronic devices.
\end{abstract}

\maketitle

Antiferromagnetic (AFM) materials form part of nearly all read sensors for magnetic recording. More recently, it has been discovered that using the AFM as the active element, instead of a ferromagnet for spintronic applications, the device can massively outperform conventional designs based on ferromagnets\cite{MacDonald2011,Gomonay2014,Jungwirth2016,BaltzRevModPhys2018,Jungwirth2018,Jungfleisch2018,Fukami2020}. Antiferromagnets are especially interesting due to their intrinsically ultrafast dynamics and higher data density. One of the most promising materials for these devices is Mn$_2$Au due to its high N\'eel temperature \cite{Barthem2013}, low in-plane anisotropy \cite{sapozhnik2018direct}, and high electrical conductivity \cite{Behovits2023}.
It has been also predicted that in  Mn$_2$Au and a similar material CuMnAs  a current induced spin-orbit-torque is able to switch the sublattice magnetic orientation~\cite{MacDonald2011,Gomonay2014}. Recently this has been demonstrated experimentally for CuMnAs~\cite{Wadley2016,Olejnk2017,Olejnk2018,MatallaWagnerPRA2019,janda2020magneto} and Mn$_2$Au~\cite{MeinertPRApp2018}, and other AFM materials~\cite{ChengPRL2020,Hajiri2019,MeinartPRR2020,Shi2020}.
In electrically driven switching of antiferromagnets the necessary current densities are very high, and with the resultant Joule heating thermal effects become especially important \cite{MeinertPRApp2018,MeinartPRR2020,2107.13859, ChiangPRL2019, Bodnar2018}. Thus, it is essential to understand the role of thermal spin fluctuations and temperature dependent magnetic properties of AFM materials.

A realistic account of the thermal effects is especially important when modelling spin-orbit-torque magnetisation dynamics and switching in nanoscale systems, which can be done with atomistic \cite{Evans2014,velez2025chaotic, RoyPRB2016, Otxoa2020, Otxoa2021, lee2024predicted,otxoa2023tailoring,rama2022temperature},  or micromagnetic~\cite{Atxitia2012,Nieves2015,PuliafitoPRB2019,SanchezTejerina2020} simulations. For the latter effective models (where thermodynamic averages of the sublattice spin order are used), the knowledge of the temperature scaling of the magnetic anisotropy, the micromagnetic exchange stiffness, and the domain wall (DW) width become crucial for a realistic modelling of electrically-induced spin-orbit-torque magnetic switching \cite{Hirst2022}.

\begin{figure}[!tb]
\centering
\includegraphics[width=15cm]{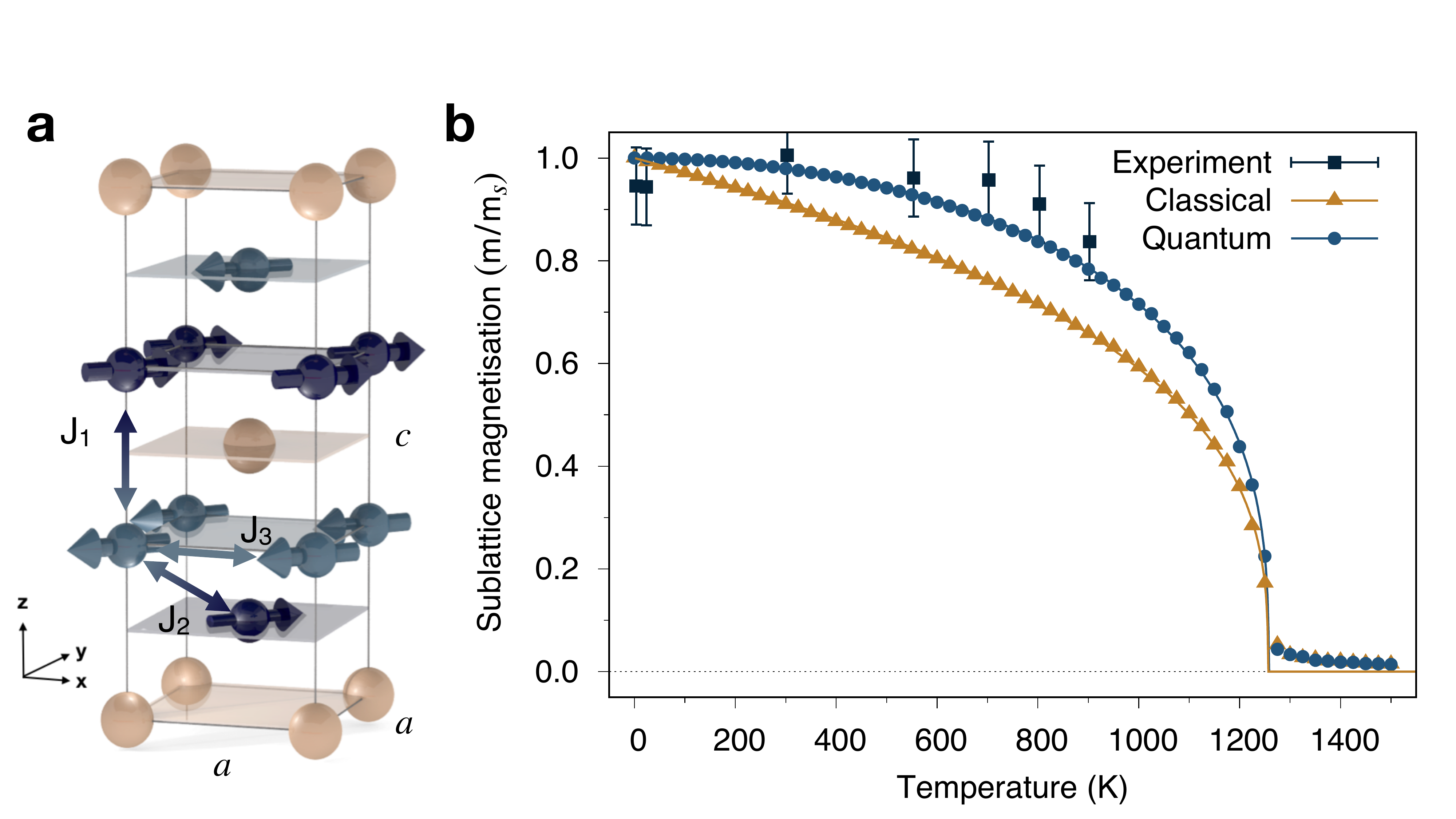}
\caption{(a) Crystal structure of Mn$_2$Au with antiparallel Mn moments (grey, purple spheres) layered between single Au sheets (gold spheres). (b) Sublattice magnetization of Mn$_2$Au as a function of temperature. The gold curve indicates the results of the direct simulations using the classical Langevin hermostat. Taking into account the quantum correction to the magnetisation versus temperature curve with Bloch exponent $\alpha=2$ (Eq. \eqref{eq:curie-bloch-law-rescaled}) to rescale the thermostat temperature allows the simulation to better match the experimental results \cite{Barthem2013} (blue curve). The simulated volume is (10 nm)$^3$. }
\label{fig:crystal}
\end{figure}

In this article, we present atomistic spin dynamics (ASD) simulations of the intricate temperature dependent magnetic properties of Mn$_2$Au antiferromagnet, guided by previous \textit{ab initio} calculations \cite{KhmelevskyiAPL2008,ShickPRB2010}. The crystal structure of Mn$_2$Au is shown in Fig. \ref{fig:crystal}. The macroscopic spin Hamiltonian, introduced for this material previously by Shick \textit{et al.} in \cite{ShickPRB2010} using a tetragonal lattice symmetry, is
\begin{equation}
    E(\theta , \phi) = K_{2\perp} \sin^2 \theta +K_{4\perp} \sin^4 \theta +K_{4\parallel } \sin^4 \theta \cos 4\phi .
    \label{eq:schick_anis}
\end{equation}
Here $\theta$ and $\phi$ are the polar and azimuthal magnetisation angles in spherical system of coordinates defined by the crystalline $z$-axis, respectively, see sketch in Fig.\ref{fig:crystal} a. The first two terms in Eq.~\eqref{eq:schick_anis} comprise uniaxial components written as a power law expansion in $\sin \theta$ for the anisotropy in second order ($K_{2\perp}<0$) and fourth order ($K_{4\perp}>0$), and the last term defines the in-plane azimuthal (in-plane rotational) anisotropy of fourth order ($K_{4\parallel}$). It is important to note that Shick \textit{et al.} \cite{ShickPRB2010} acknowledge that the uniaxial term arises from a two-ion anisotropy. Thus, the approach derived for understanding temperature-dependent parameters in the binary alloy FePt~\cite{Mryasov2005}
is  more appropriate then the direct use of the Hamiltonian-based Eq.~\eqref{eq:schick_anis} in ASD. Nonetheless, for an athermal approach Eq.~\eqref{eq:schick_anis} is perfectly valid: however, introduction of the two-ion representation is required for calculation of the temperature dependence of the macroscopic anisotropy in atomistic simulations.

Here we propose a modified atomistic Hamiltonian reflecting the exchange origin of the anisotropy. Taking into account the nature of the two-ion anisotropy in Shick \textit{et al.} \cite{ShickPRB2010}, we modify the isotropic exchange to include the anisotropic exchange coming from the $K_2$ uniaxial anisotropy. Using the modified Hamiltonian we first investigate the temperature dependence of the macroscopic anisotropies and demonstrate that they obey the expected scaling laws Zener-Akulov-from Callen-Callen (ZACC) theory~\cite{ZenerPR1954,AkulovZP1936,Callen1966TheLaw}: an important result allowing the use of temperature dependence of anisotropy for macroscopic models and the comparison with experiments.
Further, independent calculations of the DW width validate the temperature dependence of the micromagnetic exchange stiffness simulated using constrained Monte Carlo (CMC) modelling.
The results provide an  analytical expression for the DW profile with increasing temperature from low temperature up to the thermal breakdown of the $4^{\text{th}}$ order anisotropy at $\approx 800$~K. Our findings strengthen our understanding of Mn$_2$Au and open the door to an accurate theoretical exploration that can pave the way to all-AFM spintronic applications. Our demonstration of the parameterisation of all orders of anisotropy in terms of the appropriate ZACC scaling law will find application in the calculation of temperature dependent anisotropies for micromagnetic models.

The choice of our atomistic Hamiltonian is stimulated by the macroscopic spin Hamiltonian, introduced by Shick \textit{et al.} \cite{ShickPRB2010}. However unlike the macroscopic formulation, the atomistic Hamiltonian is written in terms of orthogonal functions that are better suited to produce the experimentally measured temperature-dependent properties~\cite{collings2023generalized}. It also contains the two-ion anisotropy contribution typical for bimetallic systems with heavy metal compositions (e.g., Mn$_2$Au, IrMn \cite{ShickPRB2010}, and FePt \cite{Mryasov2005}), needed to predict the physically correct temperature dependence of anisotropy. We describe the construction of the exchange tensor in the Supplementary Material Section 1. Further, as described in Supplementary Material Section 2, we write the rotational anisotropy contributions in terms of orthogonal functions.
We thus write the atomistic spin Hamiltonian in the following form:
\begin{equation}
\label{eq:hamiltonian2}
\mathcal{H} = -\sum_{i<j} \mathbf{S}_i \mathbf{J}^T_{ij} \mathbf{S}_j - k_4 \sum_i \left(S_{i,z}^4 - \frac{_{6}}{^{7}} S_{i,z}^2\right) - k_{4r} \sum_i (S_{i,x}^4 - 6S_{i,x}^2 S_{i,y}^2 + S_{i,y}^4) .
\end{equation}
where $\mathbf{S}_i$ is the unit magnetisation vector at the $i^{th}$  atomic site, $\mathbf{J}_{ij}^T$ is the exchange tensor between $i$ and $j$ spins that now contains the two-ion anisotropy term.
Note that we use capital letters $K$ to denote macroscopic, temperature-dependent variables and small $k$ for the temperature-\textit{independent} variables used in the atomistic Hamiltonian in Eq. \eqref{eq:hamiltonian2}. Thus, in Eq. (\ref{eq:hamiltonian2}), $k_4$ and $k_{4r}$ are the atomic--rather than macro--spin anisotropy terms used in Eq. (\ref{eq:schick_anis}). The values of constants can be found in Table \ref{tab:Hamiltonian_constants}.

Using ASD simulations based on the stochastic Landau-Lifshitz-Gilbert (LLG) equation \cite{Evans2014}, we first calculate the temperature-dependent equilibrium sublattice magnetisation $m_e(T)$, normalized to the value at  $T=0$ K (Fig. \ref{fig:crystal}).  These "classical" simulations give the gold curve in Fig. \ref{fig:crystal} and show a lower  magnetisation compared to the experimental results in \cite{Barthem2013} (black line in Fig. \ref{fig:crystal}).
To correct for a dependence close to the quantum case, we make use of the phenomenological temperature scaling approach from Evans~$et.\; al$ \cite{Evans2015,Anders2022}:
\begin{equation}
    \label{eq:curie-bloch-law-rescaled}
    m_e(T) = m_e(0)\biggl[1-\left(\frac{T_{\text{real}}}{T_N}\right)^{\alpha}\biggr]^{\beta}
\end{equation}
to rescale the simulation temperature using an additional critical exponent $\alpha = 2$, extracted from experiment in Ref. \cite{Barthem2013} (blue line in Fig. \ref{fig:crystal}). Note that from now on, for all temperature-dependent results will  re-scaled temperature, and the new temperature will reflect the one corresponding to the experimental situation.

\begin{table}[!h]
\centering
\begin{tabular}{c c c c c }
Interactions & $J_{xx}$ & $J_{yy}$ & $J_{zz}$ & Unit \\
\hline
$J_1$ & -1.46923 & -1.46923 & -1.45932 & $10^{-20}$ J/link \\
$J_2$ & -1.09430 & -1.09430 & -1.08691 & $10^{-20}$ J/link \\
$J_3$ &  0.31826 &  0.31826 &  0.31826 & $10^{-20}$ J/link \\
\hline
 & Parameter & \multicolumn{2}{c}{Value} & Unit \\
\hline
&$\mu_s$         & \multicolumn{2}{c}{3.72}                     & $\mu_B$ \\
&$k_{4}$           & \multicolumn{2}{c}{$3.710 \times 10^{-25}$}     & J/atom \\
&$k_{4r}$        & \multicolumn{2}{c}{$8.0255 \times 10^{-25}$}     & J/atom \\
&$T_N$        & \multicolumn{2}{c}{$1225$}     & K\\
& \textit{a,c} & \multicolumn{2}{c} {3.328, 8.539} & $\dot{\text{A}}$ \\
\end{tabular}
\caption{Simulation parameters for the atomistic spin Hamiltonian. Isotropic exchange terms are from \cite{KhmelevskyiAPL2008} and macrospin anisotropy terms from \cite{ShickPRB2010}.
}
\label{tab:Hamiltonian_constants}
\end{table}

\begin{figure}
\centering
\includegraphics[width=0.85\textwidth]{ 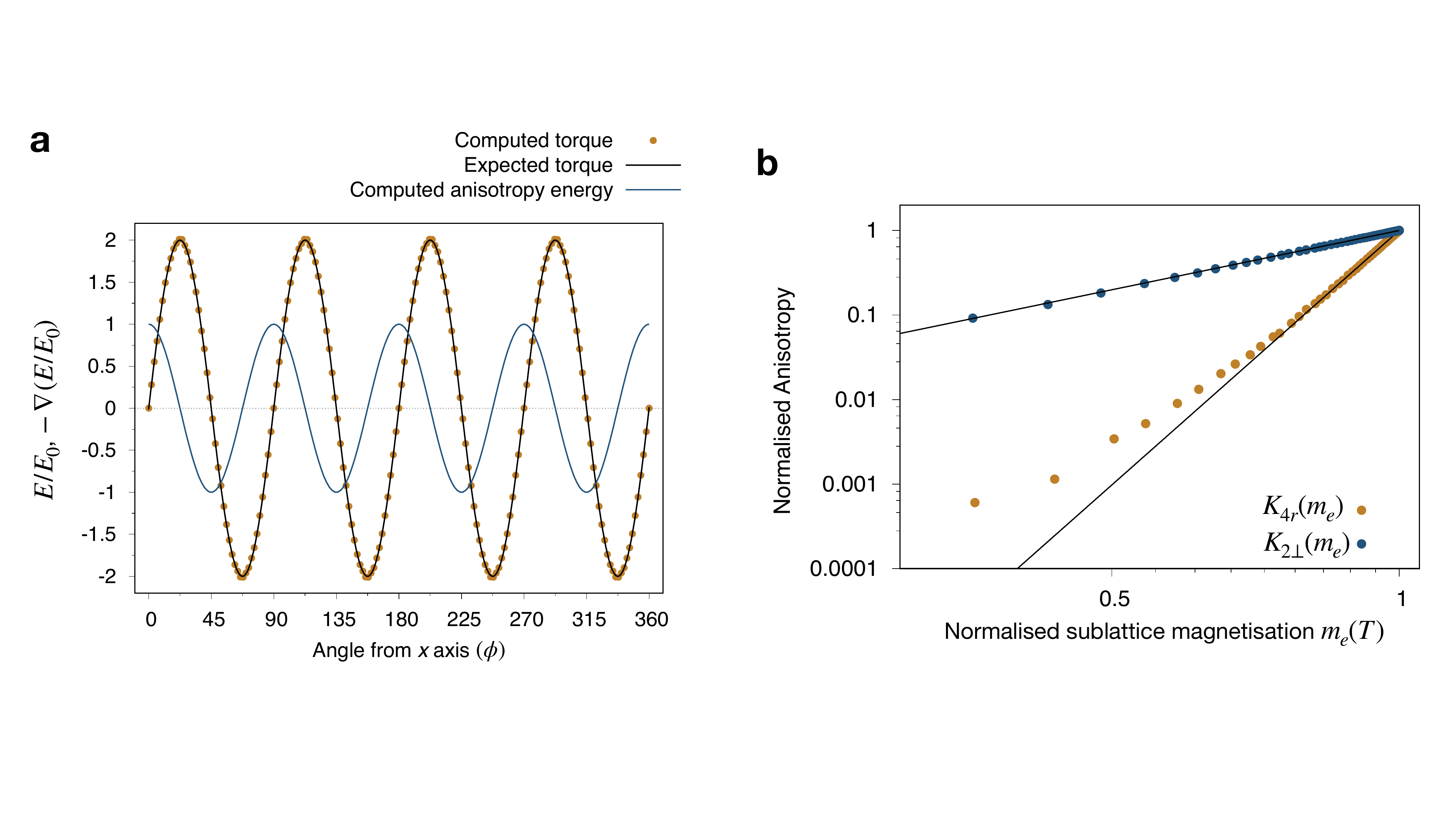}
\caption{ (a) Simulated torque curve (gold dots) and anisotropy energy (blue line) for the rotational anisotropy of Mn$_2$Au at $T = 0$ K, normalised to the rotational anisotropy constant $K_{4r}$ compared to the analytical result $-\nabla E(\theta, \phi)$ (black line).
(b) Computed temperature-dependent normalised anisotropies (points) for $K_{2\perp}$ (blue) and $K_{4\parallel}$ (gold) plotted against the normalised sublattice magnetisation $m_e$. Lines show low-temperature fits for temperature-dependent scaling of $K_{2\perp}(m_e) \propto m_e^{2.32}$--
and $K_{4r}(m_e) \propto m_e^{9.77}$.
}
\label{fig:anisotropy}
\end{figure}

Using the Hamiltonian in Eq.~\eqref{eq:hamiltonian2} for  thermal simulations, we obtain temperature-dependent anisotropies of the macroscopic Hamiltonian in Eq.~\eqref{eq:schick_anis}. For this purpose we use a CMC method~\cite{AsselinPRB2010} with an adaptive move step~\cite{Alzate-CardonaJPCM2019}. 5~000 averaging Monte Carlo (MC) steps per temperature point gives low variance in the magnetisation, though 20~000 steps was trialled to confirm the convergence. The symmetry of Eq. \eqref{eq:schick_anis} allows separation of the rotational anisotropy in the $x-y$ plane from the uniaxial contributions along the $z$-axis when the polar angle $\theta = 0$. The gradient $\bm{\tau} = -\nabla E(\theta, \phi)$,
 for the in-plane case of $\theta = 90^\circ$ gives $\tau_z |_{\theta=90^\circ} = 4 K_{4r} \sin 4\phi$ and allows the estimation of $K_{4r}(T)$.
The calculated energy (blue line) and corresponding simulated torques (gold dots) for the in-plane anisotropy are shown in Fig. \ref{fig:anisotropy}(a). It can be seen that there is a maximum for the torque at $\phi = 22.5^\circ$, so we use this point to calculate the temperature dependent $K_{4\parallel}(T)$ anisotropy, applying a quadrature rule~\cite{AsselinPRB2010} to determine the total torque.

To remove the competition between the easy-plane two-ion anisotropy and easy-axis anisotropy $k_{4}$, we chose a suitable value of $\theta = \mathrm{arccos}\left(\sqrt{15/35}\right) \approx 49.1^\circ$ where the contribution from the 4$^{\textrm{th}}$ order term is zero.
This allows us to uniquely determine the scaling of the easy-plane anisotropy.

The scaling of the anisotropies with the sublattice magnetisation $m_e$ is shown in Fig. \ref{fig:anisotropy}(b). At low temperatures the anisotropies follow the expected scaling relations  $K_{2\perp}(m_e) = K_{2}(0) m_e^{2.32}$ ($K_{2}(0) = -1.9727 \times 10^{-22}$ J/atom)  for two-ion anisotropy \cite{Rzsa2023}  and $K_{4r}(m_e) = K_{4r}(0) m_e^{9.77}$ for the 4$^{\text{th}}$ order anisotropy.   Although not presented here, the 4$^{\text{th}}$ order uniaxial anisotropy  follows the same scaling relation as the rotational term, with $K_{4}(m_e) = k_{4} m_e^{10}$.
Note that the macroscopic anisotropies are averaged over thermal fluctuations in certain magnetic volume, although expressed in energy units per atom.
Thus, the  temperature dependence of  macroscopic anisotropies closely follow the Akulov-Zener-Callen-Callen \cite{AkulovZP1936,ZenerPR1954,Callen1966TheLaw} scaling relation, frequently observed experimentally:
\begin{equation}
K(m_e) =K(T=0)\; m_e^{l(l+1)/2}
\label{eq:k4rT} ,
\end{equation}
where  $l$ is the power order of the anisotropy (e.g. 2$^{\mathrm{\text{nd}}}$, $4^{\mathrm{th}}$, etc.)

Our calculation gives the zero-temperature exchange stiffness $A(0)$ value of $2.91 \times 10^{-11}$ J/m--defined by the exchange constants used. For evaluation of exchange stiffness' temperature dependence, we use a larger system than in the above MC simulations, creating a long stripe up to 100 nm. The effective  exchange stiffness can be calculated using a different implementation of the CMC method. Here we sample the thermodynamic energy of the exchange-only Hamiltonian (Eq. \eqref{eq:hamiltonian2} without anisotropy terms) as a function of temperature with a constrained spin spiral. This is created by rotating the N\'eel vector \textit{in-plane} from parallel to antiparallel state along the $\hat{x}$ axis to construct a 180$^{\circ}$ N\'eel-type spin spiral. In the exchange limit when the magnetic texture along $\hat{y}$ and $\hat{z}$ remains uniform and the spin gradient along $\hat{x}$ is small, this method gives the exchange stiffness $A(T)$ as a function of increasing temperature.

\begin{figure}
    \centering
    \includegraphics[width=0.75\textwidth]{ 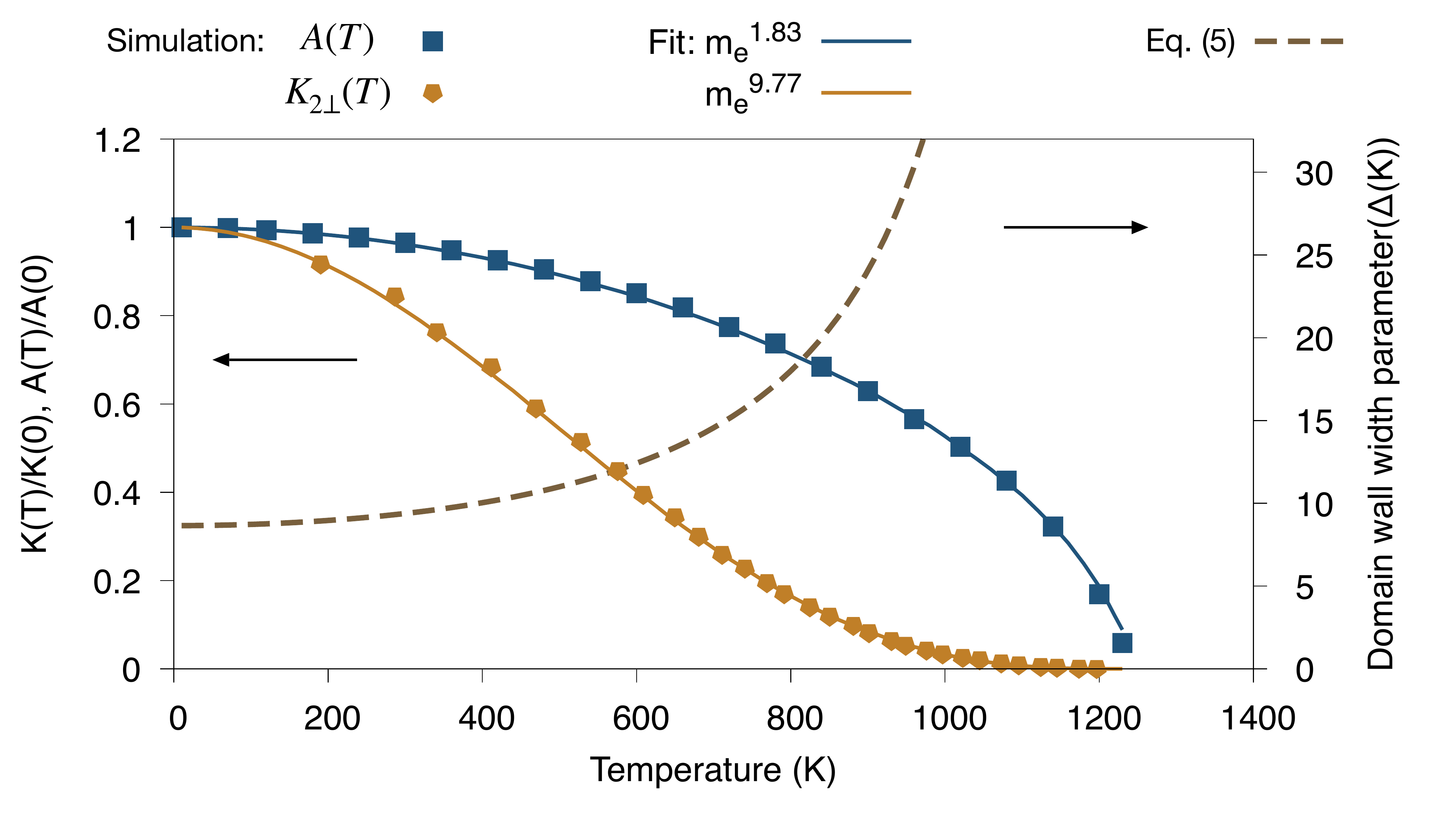}
    \caption{\textbf{Temperature dependence of exchange stiffness and anisotropy.}
    Blue and yellow lines indicate low temperature ($<200$ K) fits to the simulation data points (symbols) as a function of the equilibrium magnetisation. The brown line with no symbols corresponds to the right $y$-axis and indicates
     the DW width parameter $\Delta(T)$ for 90$^{\circ}$ DW,  calculated from the fitted scaling law for $K_{4\parallel}$ and $A(T)$ using Eq. \eqref{eq:delta-0}.
    }

    \label{fig:exchange}
\end{figure}

Fig. \ref{fig:exchange} shows  $K_{2\perp}$ anisotropy constant and exchange stiffness as a function of temperature, normalised to their zero-temperature values. A fit at low temperatures ($T< 200$ K)  gives  the exchange stiffness scaling with magnetisation of $m_e(T)^{1.68}$, agreeing with the value given in Ref. \cite{Rzsa2023}. At the same time, the temperature dependence of $A$ gives a good fit with the scaling law $m_e(T)^{1.83}$ in the whole temperature range.
The departure from the mean-field exponent 2 is in agreement with the analytical results of Ref. \cite{Rzsa2023} where it was attributed to the influence of the two-ion anisotropy.

\begin{figure}
\centering
\includegraphics[width=0.75\textwidth]{ 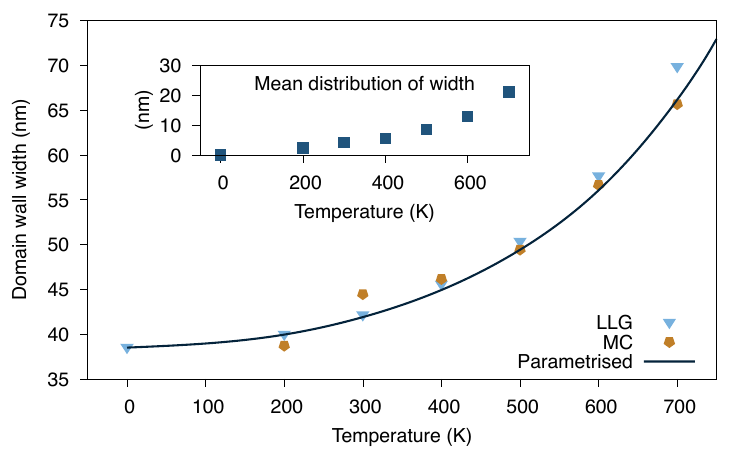}
\caption{\textbf{Temperature dependence of the average DW width, $\boldsymbol{\delta_{\mathrm{\textbf{dw}}}}$, for a 90$^{\circ}$ DW}.  Inset: mean of the distribution of the fit widths from thermal fluctuations for the damping constant $\lambda = 0.001$. }
\label{fig:tempdw}
\end{figure}

Another important characteristic is the DW width parameter $\Delta(T) $ for a 90$^\circ$ DW.
Combining our calculations for the temperature-dependent anisotropy and exchange stiffness gives the DW width parameter as a function of temperature \cite{Lendinez2026}, presented in Fig. \ref{fig:tempdw} and denoted as MC simulations.

At 0 K, the DW width $\Delta(0)$ for 90$^{\circ}$ DW is given by  \cite{Kittel, sapozhnik2018direct}
\begin{equation}
    \label{eq:delta-0}
    \Delta=\sqrt{\frac{A}{4\cdot 8K_{4\parallel}}}=8.65\; \text{nm}
\end{equation}
where $\Delta(T=0)=8.65\; \text{nm}$.
This analytic form given by Eq. (\ref{eq:delta-0}) agrees well with the numerical simulations shown later. However, the Bloch definition given by $\Delta(0)$ does not reflect the length that the magnetic texture occupies between the two domains.
Alternatively, it is possible to represent the DW width in such a way that it accounts for the slope of the in-plane azimuthal angle, $\phi$, which translates into a value $\delta_{\text{dw}}$ through the conversion $\delta_{\text{dw}}=2 \pi  \, \Delta(0)/\sqrt{2}$.

Using the previously evaluated temperature-dependent exchange stiffness and anisotropy, we can evaluate the temperature-dependent DW width $\Delta(T)$, presented in Fig. \ref{fig:exchange}.
Alternatively, the DW width can be directly simulated by MC methods or ASD. Here we use the latter. For this, we initially directed half of the Mn moments of the system along the $ \left( 1,1,0 \right)$ or along $\left( -1,-1,0 \right)$ directions, while the other half were constrained between the $\left( 1,-1,0 \right)$ or $\left( -1,1,0 \right)$ axes. For the spatially inhomogeneous region between $\left( -1,-1,0 \right)$  and $ \left( 1,1,0 \right)$ directions the DW was initialised according to the expected soliton solution given by
\begin{equation}
    \label{eq:dw-sin-temp}
  m_e(T)\sin\phi_0 = m_e(T)\sin\left(\arctan\left(\exp{\frac{x-x_0}{\Delta(T)}}\right)-\pi/4\right),
\end{equation}
where $\phi_0$ denotes the corresponding azimuthal angle of the DW boundary.
For these simulations, we increase the size of the strip length to 3000 nm to prevent finite-size effects from spin wave excitations at the boundaries.
At $T=0$ K we find $\Delta(0)
= 8.64$ nm, in very good agreement with the analytic value $\Delta(0) = 8.65$ nm.

At non-zero temperature, the system was first thermalised using a MC algorithm, then in ASD simulations we allowed to equilibrate for 20 ps, followed by 80 ps of data collection. At elevated temperatures, the DW centre experiences diffusion. We compensate for this by fitting Eq. \eqref{eq:dw-sin-temp} to each configuration snapshot, allowing the determination of the wall width parameter $\Delta(T)$.
Indeed, in view of Eqs. \eqref{eq:dw-sin-temp} and \eqref{eq:curie-bloch-law-rescaled}, a non-zero temperature will reduce the sublattice magnetisation in the domains and weaken the in-plane anisotropy-based energy landscape. Atomistically, this is caused by fluctuating moments away from the easy axes.
These are illustrated in Supplementary Material Section 3
which shows time evolution of the DW centre and the standard deviation for the time-averaged DW centres for increasing damping constants and at elevated temperatures. It was necessary to simulate for 80 ps over 10 random seeds for statistical reliability.
The consequence is that for the small system size used here even small thermal fluctuations exert a strong torque at the DW centre --where the rotational anisotropy torque is maximum--and generate large in-plane fluctuations of the atomic magnetisation \cite{Hayakawa2021}.

Taking the fitted values from Fig. \ref{fig:exchange} and substituting into Eq. \eqref{eq:dw-sin-temp}, the DW width parameter scales with the temperature-dependent magnetisation
as:

\begin{equation}
    \label{eq:mag-param}
    \Delta(T) = \Delta(0)\sqrt{\frac{m_e(T)^{1.83}}{m_e(T)^{9.77}}} \approx \Delta (0) m_e(T)^{-4}.
\end{equation}

For a more realistic functional dependence of the DW width parameter on temperature and not the magnetisation, the modified Curie-Bloch law  (Eq. \eqref{eq:curie-bloch-law-rescaled}) can be inserted into Eq. \eqref{eq:mag-param} to give:

\begin{equation}
    \label{eq:temp-param}
    \Delta(T)=\Delta(0) \left(1-\left(\frac{T}{T_N}\right)^{2}\right)^{-1.318}
\end{equation}
where for our constants $T_N=1225$ K.

Fig. \ref{fig:tempdw} shows the (re-scaled with temperature) average DW width parameters for increasing temperature for ASD and MC simulations compared to the equation parametrised from the exchange and anisotropy scaling with temperature (Eq. \eqref{eq:temp-param}). We also investigated thermal diffusion of the DW centre for different damping parameters and we saw no significant difference in the average DW width, both  on MC simulations and ASD simulations with varying damping parameters. However, again thermal activation of the DW boundary causes oscillations to the soliton's instantaneous width profile, giving a distribution of values which increases with temperature but decreases with the increase of damping. The averaged data agrees well with the parametrised calculation, up to the temperature threshold where thermal activation can overcome the in-plane anisotropy barrier ($\approx 800$ K).
Comparison with the experiment by Sapozhnik et al.\cite{sapozhnik2018direct} should take into account the re-scaled value $\delta_{\mathrm{dw}}$ which we evaluated as 38.4 nm at $T=0$ K and ca. 42 nm at room temperature that compares favourably with the range of experimentally reported values.

In summary, we have analysed the temperature-dependent properties of Mn$_2$Au using ASD and CMC simulations using a robust first principles Hamiltonian.
We first modified a magnetic spin Hamiltonian reflecting the origin of the dominant 2-ion anisotropy demonstrated in \textit{ab initio} calculations in Ref. \cite{ShickPRB2010}. The temperature scaling for macroscopic anisotropy and exchange stiffness, defined by the scaling law with magnetisation is traditionally difficult to calculate for antiferromagnets. We have calculated them for Mn$_2$Au using a CMC method. This method is confirmed by MC and LLG finite-temperature simulations of 90$^{\circ}$ DWs. The temperature-dependence of anisotropies closely follow the CC law.
The calculated exchange stiffness follows the  temperature-dependence for ferromagnets, corrected by the presence of AFM interactions \cite{EvansPRBR2020, Rzsa2023}. Increasing the value of the damping parameter decreases the diffusion distance of the DW centre. These calculations can allow large-scale micromagnetic simulations with high precision. With its high N\'eel temperature, metallic properties, and intrinsic spin-orbit torque response, Mn$_2$Au represents a crucial material for next-generation spintronic devices. With research pointing towards the  critical role temperature  in metallic spintronic devices \cite{MeinertPRApp2018}, there is a growing need for accurate, finite-temperature simulations of DW and switching processes. Thus our calculations enable more accurate multi-scale modelling of Mn$_2$Au and similar materials for understanding and control of their AFM behaviour across varying thermal conditions.

\section*{Author contributions}
R.F.L.E., J.R., S.J., R.W.C., and J.C. formulated the atomistic model parameters and anisotropy expressions. R.R.-E., R.M.O., P.G. and J.R. formulated theoretically the DW profile. J.R., S.J., P.G., and R.E. performed the atomistic simulations. J.R., O.C-F., S.J., R.W.C., and R.F.L.E. prepared the manuscript. All authors contributed to the interpretation and discussion of results.

\section*{Acknowledgements}
The authors thank Tobias Wagner, Olena Gomonay and Roberto Moreno for fruitful discussion.
S.J. and K.E.-S. acknowledge funding by the German Research Foundation (DFG) Project No.\ 320163632,
Project No.\ 268565370 - TRR 173 (project B12) and Project No.\ 405553726 – CRC/TRR 270 (project B12). J.R. and P.G. acknowledge funding from the European Union’s Horizon 2020 research and innovation programme under the Marie Skłodowska-Curie International Training Network COMRAD (grant agreement No 861300). R.R.-E. acknowledges financial support from the Royal Society through the Newton International Fellowship NIF\textbackslash
R1\textbackslash
241532, "Antiferromagnetic all-spintronics computing based on emergent electrodynamics". This project made use of the \textsc{viking} cluster, a high-performance computing facility provided by the University of York. O.C.-F. and U.A. acknowledge the support from grants PID2022-137567NB-C21, PID2024-157112OB-C52 and Severo Ochoa Centres of Excellence program  (Grant CEX2024-001445-S), funded by MCIN/AEI/10.13039/501100011033.  U.A. also acknowledges the grant “ERDF A way of making Europe” and “ESF Investing in your future”. For the purpose of open access, the
authors have applied a Creative Commons Attribution (CC BY) license to
any Author Accepted Manuscript version arising from this submission.

\bibliography{library}
\
\end{document}


\author[1]{Jackson~L.~Ross}
\author[1,2]{Sarah~Jenkins}
\author[3,4]{Ricardo~Rama-Eiroa}
\author[5,6]{Paul-Iulian~Gavriloaea}
\author[1]{Jack~B.~Collings}
\author[5]{Unai~Atxitia}
\author[5]{Oksana~Chubykalo-Fesenko}
\author[2]{Karin Everschor-Sitte}
\author[7,8]{Rub\'en M.~Otxoa}
\author[1]{Roy~W.~Chantrell}
\author[1]{Richard~F.~L.~Evans}

\affil[1]{Department of Physics, University of York, York, YO10 5DD, United Kingdom}
\affil[2]{TWIST Group,
Faculty of Physics and Center for Nanointegration Duisburg-Essen (CENIDE), University of Duisburg-Essen, 47057 Duisburg, Germany
}
\affil[3]{Institute for Condensed Matter Physics and Complex Systems, School of Physics and Astronomy, The University of Edinburgh, Edinburgh, United Kingdom}
\affil[4]{Higgs Centre for Theoretical Physics, The University of Edinburgh, Edinburgh, United Kingdom}
\affil[5]{Instituto de Ciencia de Materiales de Madrid, CSIC, Cantoblanco, 28049, Madrid, Spain}
\affil[6]{National Institute of Research and Development for Technical Physics, 47 Mangeron Boulevard, 700050 Iasi, Romania}
\affil[7]{Hitachi Cambridge Laboratory, J. J. Thomson Avenue, Cambridge CB3 0HE, United Kingdom}
\affil[8]{Donostia International Physics Center, 20018 San Sebasti\'an, Spain}

\title{Temperature-dependent magnetic properties of the antiferromagnet Mn$_2$Au: Supplementary Material}

\makeatletter
\renewcommand{\@maketitle}{%
{%
\thispagestyle{empty}%
\vskip-36pt%
{\raggedright\sffamily\bfseries\fontsize{20}{25}\selectfont \@title\par}%
\vskip10pt
{\raggedright\sffamily\fontsize{12}{16}\selectfont \@author\par}%
\vskip18pt%
}%
}
\makeatother

\maketitle

\section{Two-ion anisotropy included in the exchange tensor}

The dominant anisotropy is the two-ion, $2^{\text{nd}}$ order, uniaxial term with a negative component, giving a large in-plane preference for the magnetisation. The two-ion contribution occurs from large spin-orbit coupling in the Au layers enhancing the $3d-5d$ exchange splitting \cite{ShickPRB2010}:

\begin{equation}
    E=-k^{5d}_{2\perp}\chi^2 \sum_{\alpha,ij}J^{i,\alpha}_{3d-5d}J^{j,\alpha}_{3d-5d} M^{i,\alpha}_{3d,z}M^{j,\alpha}_{3d,z} \nonumber
\end{equation}
$\chi$ is the local spin susceptibility and $k^{5d}_{2\perp}$ is proportional to the spin-orbit coupling parameter $\xi^2_{5d}$, determined through self-consistent second-variational procedure. This suggests that the two-ion contribution is proportional to the Au and Mn bonds. We have used this proportionality and the exchange splitting $J_{3d-5d}M_{3d,z}$ to describe the calculated anisotropy constant $K_{2\perp}$ as a tensorial correction to the exchange constants \cite{Mryasov2005}.

This is done by relative weight of the exchange constants $J_1$ and $J_2$. First, the easy plane anisotropy is given by $K_{2\perp} = 2.44$ meV/f.u. which becomes $ -1.95444 \times 10^{-22}$ J/atom. Removing the $2^{\text{nd}}$ order component added to the $4^{\text{th}}$ order term $k_{2\perp} = -1.95444\times 10^{-22} - 1.60218\times 10^{-24}*40/35 = -1.9727506\times 10^{-22}$ J / atom. Note this value is already the one shown in Table 1 of the Main Text. Then, the relative exchange contributions determine the two-ion distribution: the total isotropic antiferromagnetic exchange is $-(4*1.0942959 + 1*1.469234)\times 10^{-20} = -5.8464176 \times 10^{-20}$ J/atom.
The fractional contributions to the exchange are:
\begin{align*}
J_1/J_{AFM} = \epsilon_{1} &= 1.0942959 / 5.8464176 = 0.18717374893 \\
J_2/J_{AFM} = \epsilon_{2} &= 1.4692340 / 5.8464176 = 0.25130500428
\end{align*}
The contribution of the anisotropy to each exchange value is then
\begin{align*}
k_{2\perp}J_1/J_{AFM} = 0.18717374893 * 1.9727506 \times 10^{-22} = 0.00369247125 \times 10^{-20} \; \text{J/atom} \\
k_{2\perp}J_2/J_{AFM} = 0.25130500428 * 1.9727506 \times 10^{-22} = 0.00495762097 \times 10^{-20} \; \text{J/atom}
\end{align*}
An exchange tensor incorporates the isotropic and 2-ion exchange contributions $\mathbf{J}_{ij}^T$, with $J_{xx}, J_{yy}, J_{zz}+ \epsilon_{j} k_{2\perp}$ for inter-sublattice interactions as an addition to the usual isotropic terms \cite{EvansPRBR2020}. Note that the anisotropic contribution values $< 10^{-26}$ can be truncated due to the time-step size.

We make two notes regarding the two-ion anisotropy composition above. First, for a more complete description of the effect of the two-site polarisation on exchange, one would have to redo the \textit{ab initio} calculations of Schick \textit{et al.} with site-resolved contributions for exchange and construct the spin Hamiltonian from those data. This would include the effect of the two-ion source on the $4^{\text{th}}$ order uniaxial and in-plane anisotropies. This effect, however, is minimal compared to the dominant $2^{\text{nd}}$ order anisotropy. Second, Shick \textit{et al.} calculate element specific contributions to the anisotropy orders. With minimal spin-orbit coupling on the Mn terms, it is likely that only the Au contribution is enhanced by the Mn exchange splitting. Thus, the $2^{\text{nd}}$ order anisotropy is likely a combination of single and two-ion contributions, though the two-ion contribution (-2.72 meV/f.u.) is larger than the single-ion (0.28 meV/f.u.).

\section{Cartesian description of Hamiltonian}

Substituting into the energy density expression $E=K_{4\parallel} \sin^4 \theta \cos 4\phi$ using $\cos 4\phi = 8 \cos^4 \phi - 8\cos^2 \phi +1$ gives

\begin{align*}
    E =& K_{4\parallel} \sin^4 \theta\left(8 \cos^4 \phi - 8\cos^2 \phi +1\right) \\
    E =& K_{4\parallel} \left[8 \sin^4 \theta \cos^4 \phi - 8 \sin^4 \theta \cos^2 \phi + \sin^4 \theta \right]. \nonumber
\end{align*}
The polar coordinate substitution uses $\sin^2 \theta = 1-\cos^2 \theta$, giving:

\begin{align*}
    E =& K_{4\parallel} \left[8 \sin^4 \theta \cos^4 \phi - 8 (1-\cos^2 \theta) \sin^2 \theta \cos^2 \phi + (1-\cos^2 \theta)^2\right] \\\nonumber
    E =& K_{4\parallel} \left[8 \sin^4 \theta \cos^4 \phi - 8 (1-\cos^2 \theta) \sin^2 \theta \cos^2 \phi + 1 - 2\cos^2 \theta +\cos^4 \theta\right]. \nonumber
\end{align*}
Substituting for $S_z = \cos \theta$, $S_x = \sin \theta \cos \phi$ then gives
\begin{align*}
    E =& K_{4\parallel} \left[8 S_x^4 - 8S_x^2(1-S_z^2) +1 - 2S_z^2+S_z^4\right] \\\nonumber
    E =& K_{4\parallel} \left[1+ 8 S_x^4 - 8S_x^2 + 8S_x^2S_z^2 - 2S_z^2+S_z^4\right]. \nonumber
\end{align*}

Equivalently, the energy can be expressed in terms of $S_y$ through the substitution $S_z^2 = 1 - S_x^2 -S_y^2$:
\begin{align*}
    E =& K_{4\parallel} \left[1+ 8 S_x^4 - 8S_x^2 + 8S_x^2(1-S_x^2-S_y^2) - 2(1-S_x^2-S_y^2)+(1-S_x^2-S_y^2)^2\right]. \\ \nonumber
    E =& K_{4\parallel} \left[1+ 8 S_x^4 - 8S_x^2 + 8S_x^2-8S_x^4-8S_x^2S_y^2 - 2+2S_x^2+2S_y^2-2S_x^2-2S_y^2+2S_x^2S_y^2+S_x^4+S_y^4\right]. \nonumber
\end{align*}
which reduces to give
\begin{equation}
    E= k_{4r} (S_x^4 - 6S_x^2 S_y^2 + S_y^4)
    \label{eq:4th-rot}
\end{equation}
where $k_{4r} =1/2 K_{4\parallel}$ is the atomistic, rather than macroscopic, constant.

The $4^{\text{th}}$ order uniaxial term follows a straightforward transformation to Cartesian coordinates using the identity $\sin^2\theta = 1 - \cos^2 \theta$ and $S_z = \cos\theta$:

\begin{align*}
    E=& K_{4\perp}(1-2\cos^2\theta+\cos^4\theta \\
    E=& K_{4\perp}(1-2S_z^2+S_z^4)
\end{align*}
Orthogonality is maintained by adding a small $2^{\text{nd}}$ order contribution to match the spherical harmonic form in \cite{collings2023generalized}. This shifts that anisotropy order into the form:

\begin{equation}
    E=k_{4}(S_z^4-\frac{6}{7}S_z^2)
    \label{eq:4th-uni}
\end{equation}
where $k_{4}=1/2 K_{4\perp}$ and the constant value has been removed because it does not contribute to the energy landscape of the Hamiltonian. The contribution $40/35 k_{4} S_z^2$ added to the $4^{\text{th}}$ order term must be subtracted from the $2^{\text{nd}}$ order term.
\newpage

\section{Domain wall fitting at elevated temperature}

\begin{figure}[htb!]

    \centering
    \includegraphics[width=0.85\textwidth]{ 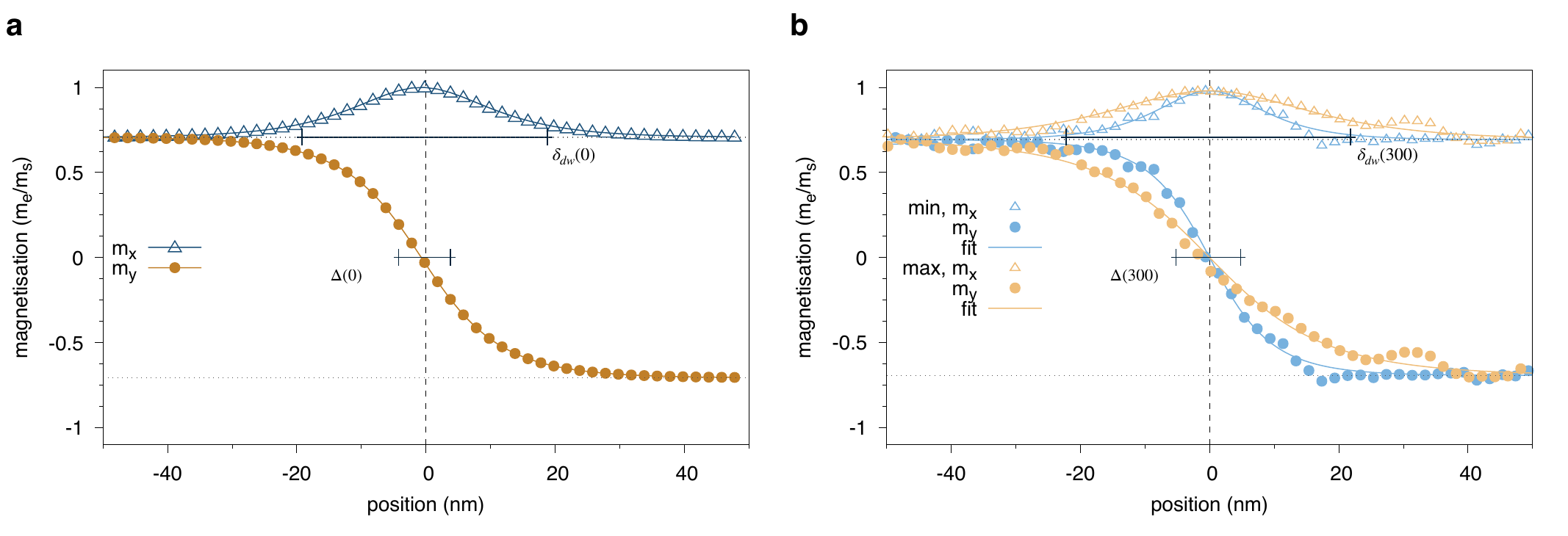}
    \caption{ \textbf{Domain wall (DW) sublattice magnetisation component profile for a) 0 K, b) 300 K.} (Light blue): instantaneous minimum fit width and points. (Gold): instantaneous maximum fit width and points.}
    \label{fig:dw-plot}
\end{figure}

\begin{figure}[htb!]
  \centering
   \includegraphics[width=0.5\textwidth]{ 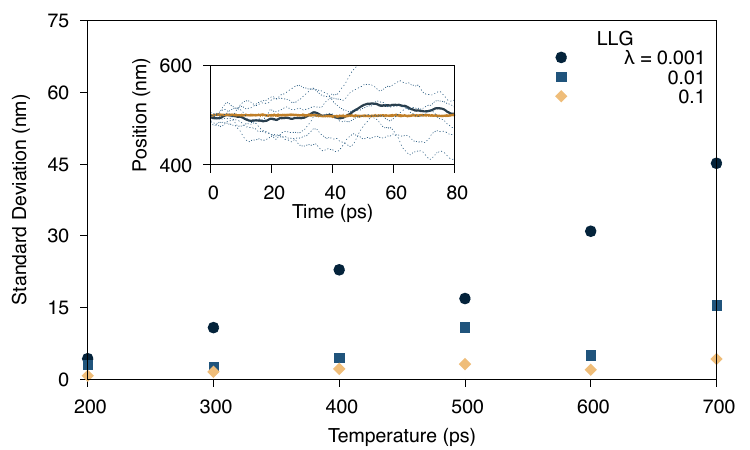}
   \caption{\textbf{DW random motion under thermal activation.} Main: Standard deviation of the time-averaged positions for different random seeds for different damping parameters. At elevated temperature the DW are highly diffusive but statistically reasonable values for the DW width was obtained by integrating for 80 pS over 10 random seeds.
   Inset: Fit DW position at 300 K for LLG random seeds with damping parameter $\lambda = 0.001$ and MC simulation. (Blue line): averaged LLG position. (Gold line): MC position. (Blue dashed): LLG random seeds.}
   \label{fig:dynamic-position}
\end{figure}

\bibliography{library}